# Decoding Stock Market Behavior with the Topological Quantum Computer

**Ovidiu Racorean**

Applied Mathematics in Finance Dept., SAV Integrated Systems

Email: rovidiu@sav-integrated.ro

**Abstract**

A surprising image of the stock market arises if the price time series of all *Dow Jones Industrial Average* stock components are represented in one chart at once. The chart evolves into a **braid representation of the stock market** by taking into account only the crossing of stocks and fixing a convention defining overcrossings and undercrossings. The braid of stocks prices has a remarkable connection with the **topological quantum computer**. Using pairs of quasi-particles, called non-abelian anyons, having their trajectories braided in time, topological quantum computer can effectively **simulate the stock market behavior encoded in the braiding of stocks**. In a typically topological **quantum computation** process the trajectories of non-abelian anyons are manipulated according to the braiding of stocks and the outcome reflects the probability of the future state of stock market. The probability depends only on the Jones polynomial of the knot formed by plat closing the quantum computation. The **Jones polynomial of the knotted stock market** acts, making a parallel with the common financial literature, in a topological quantum computation as a counterpart of a classical **technical indicator** in trading the stock market. The type of knot stock market formed is an indicator of its future tendencies.

1. Introduction

In a keynote addressed at the California Institute of Technology in May 1981 and published later in International Journal of Theoretical Physics [3], R. Feynman expressed his believe that in a near future quantum computers will be able to simulate quantum processes. Initially considered a "crazy idea", especially coming from the most unconventional physicist ever, the R. Feynman quantum computer is now close to its practical realization.

Important efforts in theoretical and engineering the construction of quantum computer have been made since then. All the efforts are well motivated considering the capacity of quantum computer to perform calculations believed to be impossible for classical computer, in an exponentially faster manner.

The quantum bits, or qubits, that quantum computer based its incredible capabilities at, are expressed as superposition states of some particles which are extremely fragile making the system to be highly sensitive to interferences with the outside world constituting the major difficulty in building a quantum computer.

This difficulty in building a quantum computer seems to be avoid by the **topological quantum computer,** a notion launched in the late 1990 and beginning of 2000, in a series of articles [4], [5], [12], [13], [14] by Freedman, Kitaev, and Preskill, just to name few of a brilliant generation of theoretical physicists.

The topological quantum computer is theoretically built around the concept of quasi-particles pairs, called non-abelian anyans, which could swap their positions clockwise or counterclockwise in a 2-dimensional plane such as their trajectories are braided in time. The topological properties of so formed braids preserve the information stored in qubits and make the system robust to outside interferences.

The picture of non-abelian anyons braiding from an initial state to a final state, in a topological quantum computation, reminisce of a recent article [1] that present stock components of *Dow Jones Industrial Average* index braiding their prices in the trading process at the stock exchange. The stocks braid is constructed simply by arranging the prices of DJIA stock components in a chart and accounting only for the crossing of stocks. To complete the image of stocks braid diagram a convention that distinguish **overcrossings and undercrossings** of stocks should be made.

Having the stock market braided it is only a simply exercise of imagination to see that manipulating pairs of non-abelian anyons by swapping their places clockwise or counterclockwise, according to the overcrossings and undercrossings of stocks braid, the topological quantum computer can easily simulate the stock market behavior.

The braiding of anyons is the typical computation process in a topological quantum computer such that the braid of stocks is actually the quantum algorithm, the quantum computation acts according with. It can be said that **trading stocks at the stock exchange is a process of writing a quantum code** and the topological quantum computer is the perfect device designed to read it, so decoding the stock market behavior.

The stock market braid is exactly replicate in quantum computation and at the end of calculation the state of the system should be read. The topological information stored in the braid is measured by quantum interference with a test anyon that is sending to braid with the system.

The outcome of the topological quantum calculation is referring at the final state of the system and expresses the probability of the stock market to end in a certain state, say bullish or bearish. The outcome probability depends only on the shape of the stocks market braid and is encoded in the Jones polynomial of the knot formed by plat closing the stocks braid.

The **Jones polynomial of the knotted stock market** acts, maintaining the proportions and exemplify with a common notion, in a topological quantum computation as a counterpart of a classical **technical indicator** in trading the stock market. So to speak the future state of the stock market is **directly dependent** of the past prices of stocks through the Johns polynomial of the knot arising by plat closing the braids formed in daily market quotation.

The present paper represents the basis of applying the topological quantum computer in practical trading applications and a lot of work is still to be done in this direction. A concrete example consisting in simulating the trading with the Fibonacci anyons will be presented in a future paper.

2. **Crossing stocks and braid of stocks diagrams**

A series of recent articles [1], [2] revealed remarkable geometrical and topological properties that the stock market is endowed with, by a simple and elegant arrangement of the prices of stocks components of a market index (*Dow Jones Industrial Average*). The most important concept that emerges from this particular arrangement of stocks is the ***crossing of stocks diagram***. The present section is reviewing some aspects related to crossing of stocks diagram and adds new insights helping the reader to get familiarized with this important notion.

The performances of a stock market as a whole are usually measured by a market index. A stock market index is a method to measure the performances of a market by a selection of the most representative stocks that are traded on the respective market. It is computed typically by a weighted average of the selected stocks prices. The levels of the market index indicate the state of the market, a market in contraction for small index levels or a market in expansion for high levels of the index.

That global view of market performances, although extremely valuable, is not offering a complete image of what happen with the stock components in the evolution of the market quotations, how the relations and correlation between them changes along the time. Some attempts to reveal the relations between market index stock components were made by T. Preis and co. and covered particular periods in the market existence such as the 2007 financial crisis, see for details [11]. This work uncovers some hidden correlations between stocks in financial crisis period and highlights the importance of accounting the influence of all stock components in the behavior of the market index.

How to analyze the price time series of market index stock components all at once? Since the easiest and familiar way of analyzing stocks prices is the **chart**, it seems that the answer is to arrange the time series of all market index stock components in **one chart**.

To simplify the discussion and give examples from real stock market conditions, and also, to celebrate its long existence, the index of interest is choosing to be the *Dow Jones Industrial Average* (DJIA). A fraction of price quotations for four DJIA components, AXP, HD, WMT and PG are shown in figure 1 as daily closing prices for a period in 2013 between 5/15/2013 and 6/7/2013, along with the chart presenting the time series of all 4 stocks.

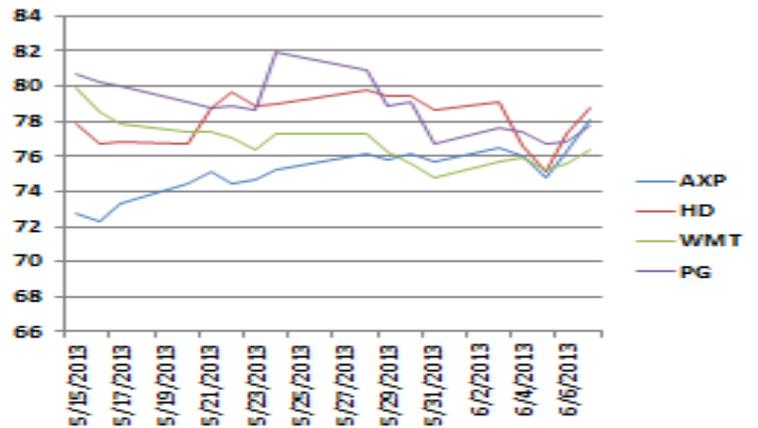

| | AXP | HD | WMT | PG |
|---|---|---|---|---|
| 6/7/2013 | 78.04 | 78.74 | 76.33 | 77.75 |
| 6/6/2013 | 76.24 | 77.26 | 75.63 | 76.82 |
| 6/5/2013 | 74.76 | 75.1 | 75.25 | 76.66 |
| 6/4/2013 | 76.06 | 76.63 | 75.94 | 77.37 |
| 6/3/2013 | 76.47 | 79.08 | 75.69 | 77.66 |
| 5/31/2013 | 75.71 | 78.66 | 74.84 | 76.76 |
| 5/30/2013 | 76.14 | 79.44 | 75.63 | 79.09 |
| 5/29/2013 | 75.83 | 79.49 | 76.23 | 78.9 |
| 5/28/2013 | 76.16 | 79.82 | 77.32 | 80.86 |
| 5/24/2013 | 75.27 | 78.99 | 77.31 | 81.88 |
| 5/23/2013 | 74.69 | 78.91 | 76.33 | 78.7 |
| 5/22/2013 | 74.44 | 79.69 | 77.03 | 78.82 |
| 5/21/2013 | 75.11 | 78.71 | 77.39 | 78.8 |
| 5/20/2013 | 74.4 | 76.76 | 77.4 | 79.09 |
| 5/17/2013 | 73.32 | 76.86 | 77.87 | 80.02 |
| 5/16/2013 | 72.23 | 76.75 | 78.5 | 80.2 |
| 5/15/2013 | 72.78 | 77.88 | 79.86 | 80.68 |

**Figure 1**. Representing all stocks prices time series in one chart at once.

It can be seen from the chart in figure 1 that trajectories of stocks prices are crossing and re-crossing each other as one stock price exceed or came under the price of a neighbor stock, in other words every time the *stocks are crossing*. The most important concept residing from this particular arrangement of the market index stock components is the **crossing of stocks.**

Following the colored stock time series in the chart above a stocks crossing diagram can be drawn. The *crossing of stocks diagram* explicitly shows the moments when the price of a stock comes over or under the price of its neighbor stock as is shown in figure 2.

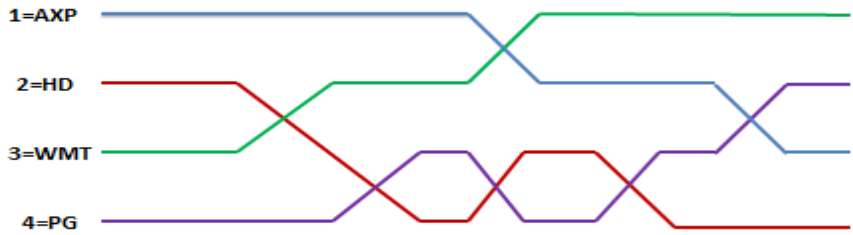

**Figure 2**. Crossing of stocks diagram, a stylized representation of the chart in figure 1.

Figure 2 is a replica of the chart in figure 1 consisting in all stocks time series showing only the crossing of stocks. Notice that it cannot be said from the crossing diagram which is the stock that have the price overcoming or under coming the price of its adjacent stock. In the chart in figure 1 it can immediately notice that in the interval from 5/20/2013 to 5/21/2013 the price of HD came over WMT and PG, even only by observing the graph of every stock.

To indicate this crucial information in the crossing of stocks diagram for every crossing of two adjacent stocks the difference between the price after and before the crossing are calculated for both stocks. The differences are taking in modulo since only the net amounts are considered, such that:

$$\Delta_{Stock\ i} = |P_{before\ crossing} - P_{after\ crossing}|, \qquad (1)$$

$$\Delta_{Stock\ i+1} = |P_{before\ crossing} - P_{after\ crossing}|, \qquad (2)$$

where P is the price of the stock. The stock with the higher difference it will come over and the stock with the smaller difference will be under in a stock crossing.

The two cases that can arise are:

- $\Delta_{Stock\ i} > \Delta_{Stock\ i+1}$ - the stock $i$ is crossing over the stock $i + 1$, in which case the stocks crossing will be called to be an **overcrossing of stocks**,
- $\Delta_{Stock\ i} < \Delta_{Stock\ i+1}$ - the stock $i$ is crossing under the stock $i + 1$, in which case it will be called to be an **undercrossing of stocks**.

The two situations in discussion for the crossing of two stocks are exemplified in the figure 3.

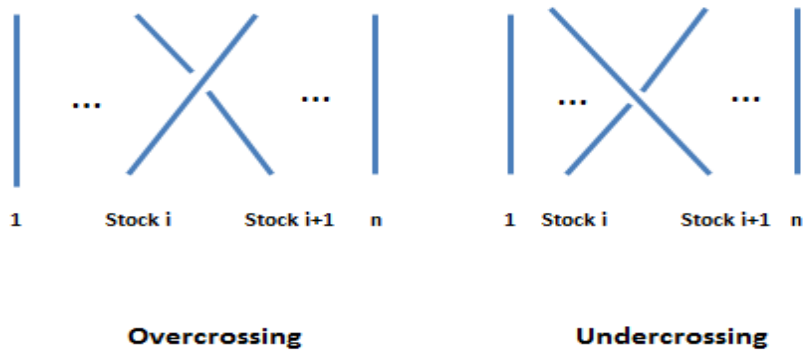

**Figure 3.** Overcrossing and undercrossing of stocks.

The explicit functionality of this schema can be shown in a simple example coming from the real market price quotations of stocks. Let's get back to the figure 1 b) and analyze the first crossing of stocks that show up at 05/21/2013 between stocks HD and WMT. Their Δ differences are:

$$\Delta_{HD} = |77{,}39 - 77{,}40| = |-0{,}01| = 0{,}01$$

$$\Delta_{WMT} = |78{,}71 - 76{,}76| = |1{,}95| = 1{,}95$$

such that $\Delta_{HD} < \Delta_{WMT}$ and it is an undercrossing between these two stocks.

Evaluating all the crossings of stocks in the diagram shown in figure 2 the new representation of the stock market (here only for 4 stock components of DJIA) is shown in figure 4.

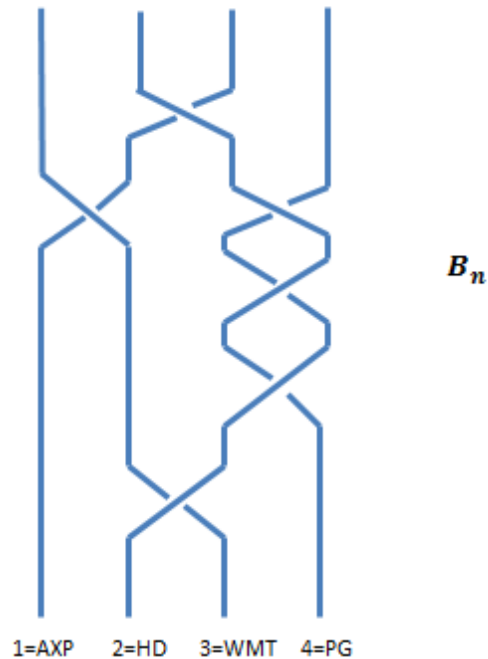

**Figure 4**. The braid representation of a fragment of stock market Dow Jones Industrial Average index consisting in 4 stocks with price quotations in the interval from 5/15/2013 to 6/7/2013.

The diagram of the crossing of stocks is ported vertically for later convenience. The stocks prices as it can be easily noticed from the figure 4 are **braided** from 5/15/2013 to 6/7/2013.

The braid that stocks form is endowed with topological properties. Although braids are known in the day by day experience since the old times only relatively recently Emil Artin uncover the powerful mathematical structure of braids in his paper written in the middle of 1920' and refine it in 1947 [22].

Details of stocks braids topological properties can be find in [1] and the interested reader is encouraged to explore it in details.

To investigate more topological properties of the braided stocks that will be needed in the next sections, the Alexander Theorem according to every close braid results in a knot is used further. The braided stocks are tied into knots and links and their topology is studied.

3. **Tying the stock market into knots and links**

There are two classical ways for closure of braids into knots:

- Trace closure;
- Plat closure.

The **trace closure of a stocks braid** has been investigated in [1]. The present paper extends the applications of topological notions to stock market by exploring the **plat closure of stocks braid** that will be central in the further sections.

Simply speaking the plat closure of a braid is obtained by gluing together at the bottom and at the top, the adjacent strands of a braid starting from the left at the braid as it can be seen in figure 5.

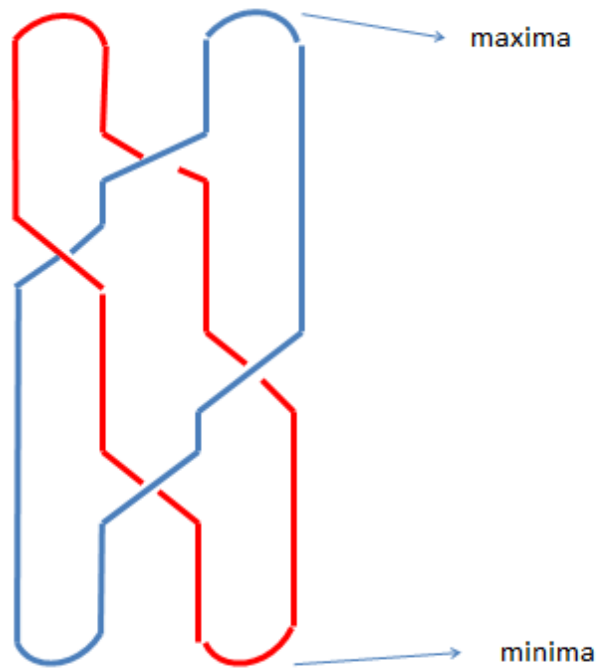

**Figure 5**. The Plat closure of the stocks braid results in a Hopf link. The **minima** and **maxima** of the closure are explicitly indicated.

It can be easily noticed that a necessary condition for Plat closure of a braid is to have 2k strands such as the resulting closure will have k maxima and k minima. It is exemplified in figure 5 that for the 4 stocks results 2 minima and 2 maxima.

The closure of the braid in figure 5 represents the well-known Hopf link and its two components were drawn in different colors to make the identification easier.

In the daily stock market trading activity many crossings of stocks appear due to price quotations, such that the braid representation changes in time. The closure of the formed braids of stocks will results in different types of knot or link. To exemplify, taken the same fraction of 4 DJIA stocks prices from 5/15/2013 to 6/5/2013 the plat closure of the formed braid is represented in figure 6.

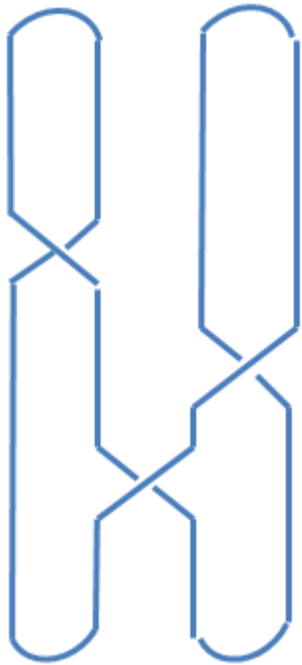

**Figure 6.** The plat closure of the   showing a different knot

In this case the knot that results from the plat closure of the stocks braid is the **unknot**.

These two simple examples shows to be easy to evaluate the knot related to the closure of the stocks braid. What if all the 30 stocks components of Dow Jones Industrial Average are involved? The stocks braid will then have 30 strands with many crossings such that the type of knot related to the closing of such a braid will prove to be extremely difficult cu find. The method to find and distinguishing between types of knot is to calculate their polynomial invariants.

The importance of distinguished between types of knot created in the stock market, meaning to differentiate between trajectories formed by stocks prices, will become obvious in the next sections.

## 4. Jones polynomial and Kauffman brackets

Prior to enter deeper in the problematic of calculating knot polynomial invariants a simply but elementary notion should be clarified.

**Writhe** ($Wr(K)$) of a knot or link diagram is the total of positive crossings minus the total number of negative crossings:

$$Wr(K) = \sum(Positive\ crossings - Negative\ crossings) \qquad (3)$$

For the sake of exemplification, the writhe for the link in figure ... having 3 overcrossings and one undercrossing is calculated as $Wr(K) = 3 - 1 = 2$.

Distinguishing between types of knot is related to calculate the polynomials that are invariants of knot diagrams. Since the discovery of Jones polynomial in the middle of 1980, knot theory flourished and many invariants were found. To make the connection with the next sections of the paper only the relevant polynomials for the ideas that will be expressed further, Jones polynomial and Bracket polynomial are discussed.

The polynomial for a knot is obtained, by applying the **skein relation** to every crossings appearing in the knot diagram. It was exemplified in [1], for the link formed in the stock market from 5/15/20133 to 6/7/2013 how the skein relation

$$\frac{1}{t}V_{K_+}(t) = tV_{K_-}(t) + \left(\sqrt{t} - \frac{1}{\sqrt{t}}\right)V_{K_0}(t), \qquad (4)$$

accompanied with the rule for the unknot

$$V_K(t) = V_O = 1,$$

discovered by V. Jones in [6] is applied to calculate the **Jones polynomial** $V_K(t)$.

The Jones polynomial is the most celebrated knot invariant and its applications are vast and spread over numerous branches of research, from quantum physics to biology, and now finance.

***Bracket polynomial*** $\langle K \rangle$ assigns a bracket to every knot diagram and its skein relations are based on the notations from the figure 7.

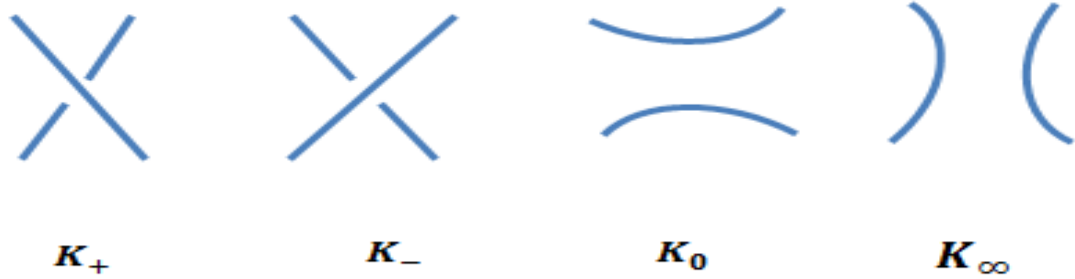

**Figure 7**. Notation that Kauffman bracket skein relation is based at.

The skein relations of the Kauffman brackets and the rule for the unknot necessary to calculate the related polynomial are:

$$\langle K_+ \rangle = A \langle K_0 \rangle + A^{-1} \langle K_\infty \rangle \tag{5}$$

$$\langle K_- \rangle = A^{-1} \langle K_0 \rangle + A \langle K_\infty \rangle \tag{6}$$

$$\langle KO \rangle = (-A^2 - A^{-2}) \langle K \rangle \tag{7}$$

The reader is referring to the work of L. Kauffman [8], [9] and [10] to see examples of calculations of bracket polynomial for some familiar knot diagrams.

There is a close connection between Jones polynomial and Bracket polynomial that will be exploited further, and is express by the relation:

$$f[K] = (-A)^{-3Wr(K)} \langle K \rangle \tag{8}$$

and

$$V_K(t) = f[K]\left(t^{-1/4}\right) \tag{9}$$

where $f[K]$ is the Kauffman invariant, but having no importance for the present paper will be ignored for now.

Renormalizing and rearranging the terms the bracket polynomial for a knot or link can be written as:

$$\langle K \rangle = (-A)^{3Wr(K)} V_K(A^4) \tag{10}$$

The relation is central in topological quantum computation, a fascinating subject that will be explored further, and also it will be seen what tying the stock markets in knots and links is good for.

### 5. Topological Quantum Computer

Since R. Feynman [3] launched the incredible idea of a quantum computer that could simulate the probabilistic world of subatomic particles, important steps forward in theoretical and engineering the construction of such remarkable device have been made. The search for a quantum computer is mainly motivated by its capacity to perform calculations believed to be impossible for classical computer. A quantum computer is also exponentially faster than a digital computer. To make a parallel remark with financial world of trading, just imagine how is like to trade stocks with a quantum computer, the high frequency trading of today is going to be the very low frequency trading of tomorrow.

The incredible capabilities of quantum computers reside in their usage of quantum bits, or qubits, instead of classical computer bits. The qubits are expressed as quantum properties of some physical particles. The superposition states of such particles are extremely fragile and that makes the system to be highly sensitive to interferences with the outside world which is the major difficulty in building a quantum computer.

To avoid this difficulty and protect the information stored in qubits, in a series of papers [4], [5], [12], [13], [14], Freedman, Kitaev, and Preskill, to name few of a brilliant generation of theoretical physicists, launched in the late 1990 and beginning of 2000, the notion of **topological quantum computer**.

Topological quantum computer is based on the remarkable simply and beautiful idea that topological properties of an object remain unchanged under deformations. Endowed with topological properties a physical system is insensitive to local perturbations and protects information from errors caused by interactions with the environment.

The "heart" of topological quantum computer consists in pairs of quasi-particles, called anyans, which could swap positions in a 2-dimensional plane. The topological properties of this quantum system appear out of the anyons trajectories in time, as figure 8 shows.

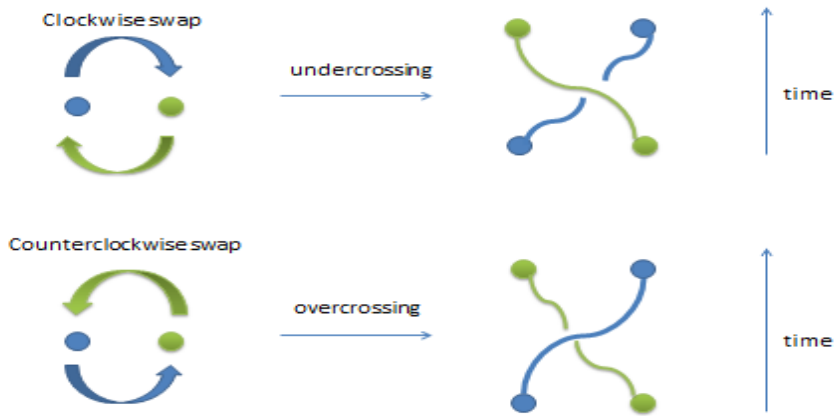

**Figure 8**. Non-abelian anyons exchanging places and their braiding trajectories. A clockwise swap is similar to an undercrossing and counterclockwise swap is reflected an overcrossing.

It can be noticed from figure 8 that a clockwise swap is reflected in an overcrossing and a counterclockwise swap in an undercrossing, of a braid representation. The two possible anyons paths are topologically distinct. To distinguish between these two possible quantum states, the topological quantum computer needs a particular type of anyons, called **non-abelian anyons**. Non-abelian anyons have the property that the order in which particles are exchange positions is important; it said that exhibit braiding statistics.

Not entering in technical details, that can be found in [], [].. it is suffice to point out here that the final state $\Psi_f$ for a quantum system consisting in a pair of non-abelian anyons $a$ and $b$, having the same initial state defined by the wave function $\Psi_i$, differs as follow:

- For a clockwise swap of positions - $\Psi_f \rightarrow M_{ab}\Psi_i$ ;
- For a counterclockwise swap - $\Psi_f \rightarrow N_{ab}\Psi_i$ .

The matrices $M_{ab}$ and $N_{ab}$ do not commute, such that:

$$M_{ab} \neq N_{ab} . \qquad (11)$$

The braided trajectories that results by manipulating the non-abelian anyons in clockwise and counterclockwise moves isolate the quantum information from errors.

Figure 9 depicted the braid formed by the trajectories of two pairs of non-abelian anyons by sequential swaps of positions between them.

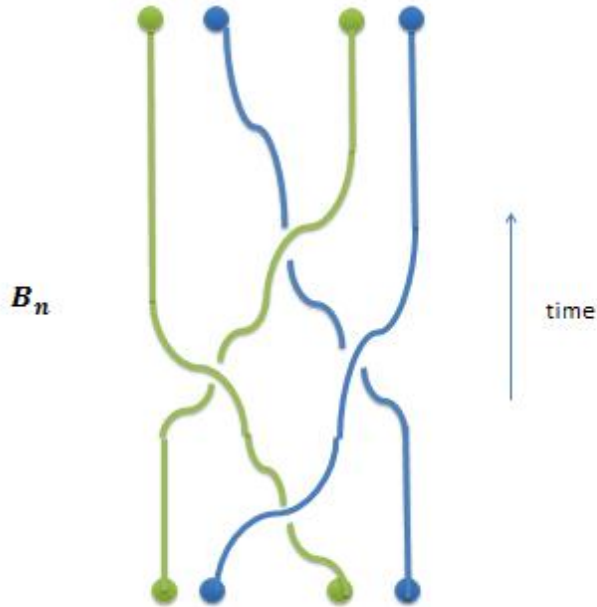

**Figure 9.** Braided trajectories of two pairs of non-abelian anyons.

The final state of the system depends on the initial state and the braid that anyon trajectories form by moving them around which other:

$$|\Psi_{final}> = B_n |\Psi_{initial}> . \qquad (12)$$

## 6. Topological Quantum Computation

The topological quantum computer, as it can easily be noticed is far away from the conventional image of a classical computer. Since the "hardware" is that different, the natural question to ask is how the "software" could look like? The hint that last section traced is that the computation with a topological quantum computer is nothing else than braiding the non-abelian anyons in a predefined series of clockwise and counterclockwise moves.

The typical topological quantum computation process consists in the following steps:

- Pairs of non-abelian anyons are created out of the vacuum and lined up in qubits of the computation;
- The anyons are manipulated by swapping position of adjacent anyons in a predefined sequence; moving around the anyons corresponds to operations performed on qubits;
- Pairs of adjacent anyons are brought together and fused to close the computation process. Measuring the final state of the system produce the outcome of the computation.

The outcome of the topological quantum computation depends only on the topology of the braided trajectories of the anyons.

Every possible sequence of anyon manipulation can be put in correspondence with a certain braid.

Figure 10 shows in detail a typical topological quantum computation operation based on two pairs of non-abelian anyons that by a series of 3-counterclockwise and 1-clockwise moves formed the braid $B_n$.

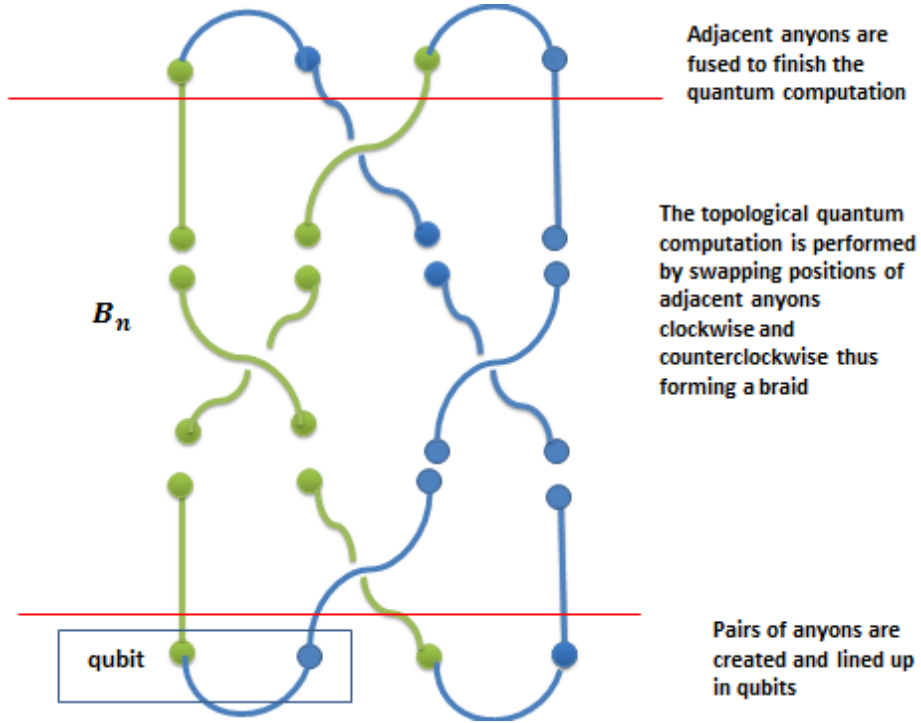

**Figure 10**. The typical topological quantum computing with braided non-abelian anyons.

Notice from figure 10 that **creation of pairs of anyons, braid their trajectories and fusing them together at the end of the computation result in a plat closure of braided anyons**. This important conclusion opens the gates to knot applications that it was seen in the latter sections are topologically more complex.

In the relation (12) the braid $B_n$ could be replaced with the plat closure to take into account the creation and fusion of the anyons such that:

$$|\Psi_{final}> = \ plat(B_n) \ |\Psi_{initial}> \tag{13}$$

The outcome of the computation typically expresses the probability to find the system in a certain final state.

The probability of finding the system in a certain final state, say 0 is:

$$prob(0) = \langle \Psi_{initial}|\Psi_{final}\rangle = \langle \Psi_{initial}|plat(B_n)|\Psi_{final}\rangle \tag{14}$$

and is directly dependent on the plat closure of the braids resulting from anyon manipulation.

### 7. Simulate the Stock Market behavior by manipulating non-abelian anyons

The braiding of stocks prices in the stock market as it was defined in the previous sections seems now more like a code that topological quantum computer is the perfect machine designed to decode it.

Remaining in the limit reflected by a fraction of 4 DJIA stock components the braiding that results in price quotations of stocks from 5/15/2013 until 6/7/2013 can be easily simulated by the topological quantum computer.

Two pairs of non-abelian anyons (two qubits) are pulling out of the vacuum state and arranged in the initial configuration in qubits. The adjacent particles are swapped clockwise or counterclockwise according to the prices of stocks moves as it can be depicted in the figure 11.

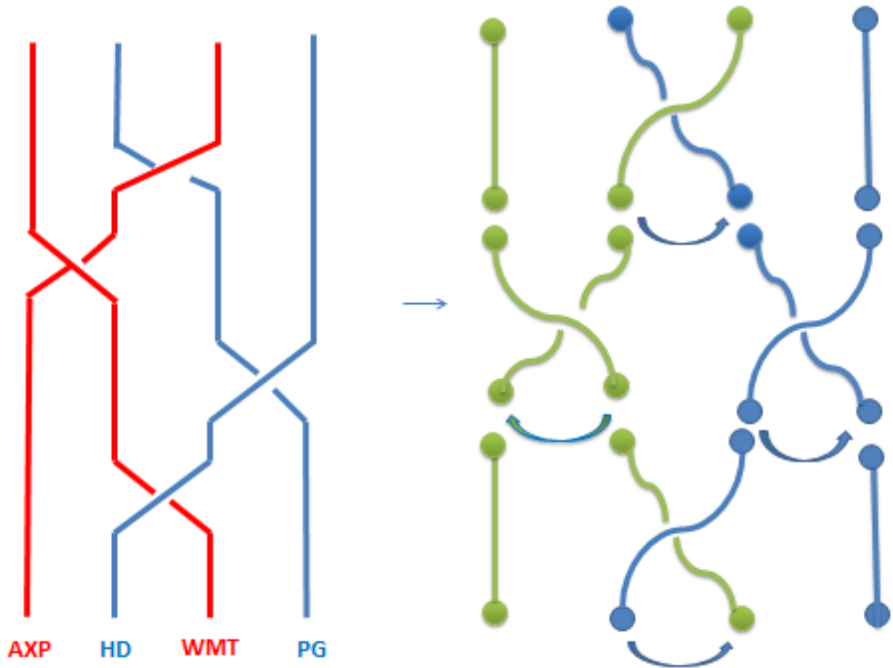

**Figure 11**. Parallel between the braid of stocks and the braided trajectories of non-abelian anyons. The arrows show how non-abelian anyons are manipulated clockwise and counterclockwise to explicitly simulate the braided stocks.

Figure 11 shows the simplified version of the stocks braid in figure 4 and the simulation realized by manipulation of non-abelian anyons in braiding trajectories.

The simulation of the stock market is actually a topological quantum computation over the qubits such as at the market close the output of this process should be the probability of the future states of the stock market.

How could this important result be read after the end of quantum computation when the initial anyons are brought together and fused?

## 8. Topological Quantum Algorithm to decoding the Stock Market Behavior

The trajectories of non-abelian anyons are manipulated to explicitly follow the stock market behavior and thus effectively replicate the market behavior.

The question that arises in is how to read the final state of the system (the final state of the stock market) at the end of computation process such that the topological information can be of any use.

Following the prescription of Freedman, Kitaev and co. in [12] one possibility is to send a test quasi-particle in the system such that its trajectory is allowed to braid with the trajectories of the first pair of non-abelian anyons, letting the quantum information being read by interference of trajectories.

The constrain that any quantum computation should be reversible add to the system consists in braid $B_n$ (simply $\sigma$), its inverse $B_n^{-1}$ (or $\sigma^{-1}$). Under these circumstances the final prescription for the topological quantum computation, noting the trajectory of the test quasi-particle with $\gamma$, takes a more complicated form that is shown in the figure 12.

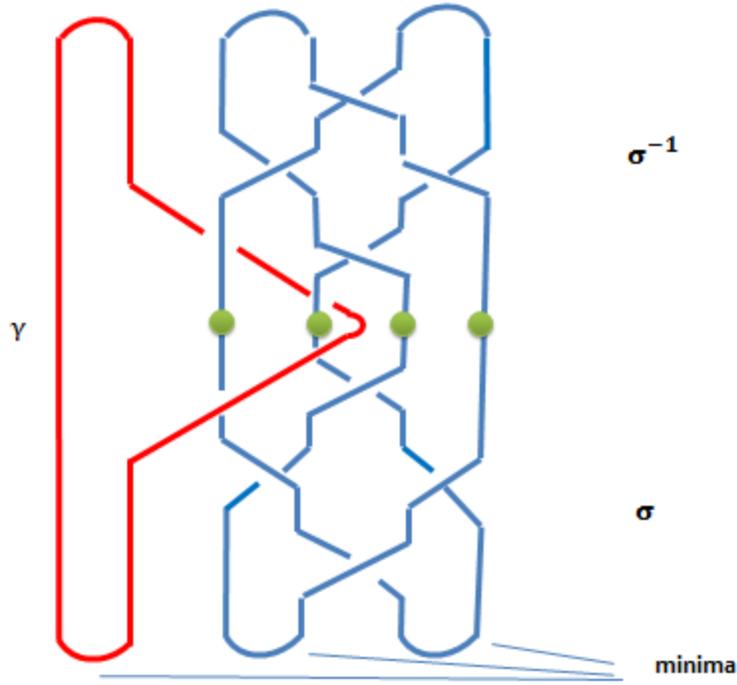

**Figure 12.** Quantum interference of the system with a test quasi-particle (in red) to read the outcome of topological quantum computation.

The braid that results in combining all the above prescriptions is:

$$Br_n = \sigma_\gamma \sigma^{-1} \qquad (15)$$

The final state of the system after applying the prescriptions above is the same as the initial state such that the outcome of the quantum computation now takes the form:

$$prob(0) = \langle \Psi_{initial} | plat(Br_n) | \Psi_{initial} \rangle = \langle \Psi_{initial} | plat(\sigma_\gamma \sigma^{-1}) | \Psi_{initial} \rangle \qquad (16)$$

If the plat closure of the braid is expressed in terms of Kauffman brackets, for the outcome of the computation the relation bellow holds:

$$\langle \Psi_{initial} | plat(\sigma_\gamma \sigma^{-1}) | \Psi_{initial} \rangle = \frac{1}{[2]_5^{n/2-1}} \langle K(\sigma_\gamma \sigma^{-1}) \rangle \qquad (17)$$

where $n$ is the number of strands in the braid representation.

The Kauffman brackets appearing in the formula (17) is known from a latter section and accounting the connection with the Jones polynomial the probability of a certain final outcome, after some calculations following [12] and [18], is expressed as:

$$prob(0) = \frac{1}{1+[2]_5^2}\left(1 + \frac{(-1)^{c(K)+Wr(K)}(-A)^{3Wr(K)}V_L(A^4)}{[2]_5^{m(K)-2}}\right). \tag{18}$$

Leaving for a future paper the details of applying Fibonacci anyons to the stock market it should be said here that in the hypothesis that this particular non-abelian anyons obeying the braiding statistics, the variable $A$ is chosen to be $A = e^{\pi i/10}$, such that the relation above becomes:

$$prob(0) = \frac{1}{1+[2]_5^2}\left(1 + \frac{(-1)^{c(K)+Wr(K)}(-A)^{3Wr(K)}V_L\left(e^{2\pi i/5}\right)}{[2]_5^{m(K)-2}}\right), \tag{19}$$

where $[2]_5 = \frac{1+\sqrt{5}}{2}$, $c(K)$ represents the number of components, $m(K)$ is the number of minima of the knot $K$ and $Wr(K)$ is the latter discussed writhe of $K$. Discussions referring to the writhe appearing in the above relation can also be found in [12], [18].

The probability of finding the system in a certain final state is **only depending on the Jones polynomial** is the final statement of the above equation.

Referring to the financial applications, the **Jones polynomial of the knotted stock market** acts, maintaining the proportions and exemplify with a common notion, in a topological quantum computation as a counterpart of a classical **technical indicator** in trading the stock market. Speaking in terms of technical trading, a certain knot formed in the stock market is an indicator of the tendencies that market will have in the future.

The future state of the stock market is **directly dependent** of the past prices of stocks through the Johns polynomial of the knot arising by plat closing the braids formed in daily market quotation. **The relevant information of the stock market behavior is encoded in the way stocks are braided.**

### 9. Final remarks

The recent article [1] presents a surprising image of stock market as braids, knots and links of stocks prices. The present paper first sections briefly reviews these mathematical notions and enriches the topological frame of stock market with other concepts meant to complete the picture that topological quantum computation will be applied at, in the last sections.

Braided of stocks arise simply by considering the price time series of all the Dow Jones Industrial stock components in one chart at once. Taking into account only the crossings of stocks in the chart and fixing a convention defining the **overcrossing** of stocks, if the price of a stock gets over the price of an adjacent stock, and **undercrossing** of stocks, if the stock price come under the price of a neighbor stock, the chart evolve in a **braid representation of the stock market**.

To prepare well in advance the terrain for the sections dedicated to topological quantum computer, the knot formation out of plat closing the braid of stocks is explain and two relevant polynomial invariants, Jones polynomial and Kauffman brackets, are explained. The remarkable relation connecting these two knot invariants is discussed.

The **topological quantum computer** is the most promising theoretical attempt in engineering a quantum computer, a device believed to be capable of calculations beyond the capacity of the ordinary computers. The core of topological quantum computer, so to speak the "hardware" consists in pairs of quasi-particles, called non-abelian anyons that could swap their positions clockwise or counterclockwise, in 2-dimensional plane such that their trajectories formed braids. It can be easily hint from this picture the connection with the stock market braids.

Manipulating the non-abelian anyons clockwise or counterclockwise according to the overcrossings or undercrossings of the braiding stocks, the **stock market behavior is exactly simulated** in the topological quantum computing process. It can be said that the braid of stock market is nothing else than the "software" for the topological quantum computer.

**Trading stocks at the stock exchange is a process of writing a quantum code** and the topological quantum computer is the perfect device designed to read it, so decoding the stock market behavior.

The end of typical topological quantum computation consists in fusing the pairs of non-abelian anyons together, a process that results in plat closure of the braided trajectories of anyons. The outcome of the topological quantum calculation is referring at the final state of the system and expresses the probability of the stock market to end in a certain state, say bullish or bearish. The outcome probability depends only on the shape of the stocks market braid and is encoded in the Jones polynomial of the knot formed by plat closing the stocks braid.

The **Jones polynomial of the knotted stock market** acts, maintaining the proportions and exemplify with a common notion, in a topological quantum computation as a counterpart of a classical **technical indicator** in trading the stock market. So to speak the future state of the stock market is **directly dependent** of the past prices of stocks through the Johns polynomial of braids formed in daily market quotation.

The present paper puts the basis of applications of topological quantum computer in practical financial issues and a lot of work is still to be done in this direction. A concrete example consisting in simulating the trading of stocks with the Fibonacci anyons will be presented in a future paper.

**References**

[1] O. Racorean, Braided and Knotted Stocks in the Stock Market: Anticipating the flash crashes, http://arxiv.org/abs/1404.6637, 2014.

[2] O. Racorean, Crossing of Stocks and the Positive Grassmannian I : The Geometry behind Stock Market, http://arxiv.org/abs/1402.1281, 2014.

[3] R. Feynman, Simulating physics with computers, Int. J. Theor. Phys.21(1982), 467-488.

[4] A. Yu. Kitaev, "Fault-tolerant quantum computation by anyons," Annals Phys. 303, 2-30 (2003), arXiv: quant-ph/9707021.

[5] R. W. Ogburn and H. Preskill, Topological quantum computation, Lect. Notes in Comp. Sci. 1509, 341-356, (1999).

[6] V.F.R. Jones, A polynomial invariant for knots via von Neumann algebras, Bull. Amer. Math. Soc. 12 103‑112,1985.

[7] V.F.R. Jones, The Jones Polynomial, 2005, http://math.berkeley.edu/~vfr/jones.pdf.

[8] L.H. Kauffman, State models and the Jones polynomial, *Topology* , **26** (1987) pp. 395–407.

[9] L.H. Kauffman, An invariant of regular isotopy, *Trans. Amer. Math. Soc.* , **318** : 2 (1990) pp. 417–471.

[10] L. H. Kauffman and S. J. Lomonaco Jr., Topological quantum computing and the Jones polynomial, 2006, quant-ph/0605004.

[11] Kenett DY, Preis T, Gur-Gerschgoren G, Ben-Jacob E. Quantifying meta-correlations in financial markets. EPL 99, 38001, (2012).

[12] M. H. Freedman, A. Kitaev, M. J. Larsen and Z. Wang, Topological Quantum Computation, Bull. Amer. Math. Soc. 40, 31 (2003).

[13] M. H. Freedman, A. Kitaev, Z. Wang, Simulation of topological field theories by quantum computers, Commun. Math. Phys., 227, 587-603, (2002), quant-ph/0001071.

[14] J. Preskill, Topological computing for beginners, (slide presentation), Lecture Notes for Chapter 9 - Physics 219 - Quantum Computation, http://www.iqi.caltech.edu/preskill/ph219.

[15] N. E. Bonesteel, L. Hormozi, G. Zikos and S. H. Simon, Braid topologies for quantum computation, quant-ph/0505665.

[16] D. Aharonov, V. Jones, Z. Landau, A polynomial quantum algorithm for approximating the Jones polynomial, quant-ph/0511096.

[17] C. Nayak, S. H. Simon, A. Stern, M. Freedman and S. Das Sarma, Non-Abelian Anyons and Topological Quantum Computation, Rev. Mod. Phys. 80, 1083 (2008).

[18] M. Bordewich, M. Freedman, L. Lovasz, D. Welsh, Approximate Counting and Quantum, Computing, Combinatorics, Probability and Computing 14, 737-754, 2005, http://arxiv.org/abs/0908.2122.

[19] Fama E., The Behavior of Stock Market Prices, J. Business 38, pp. 34-105, 1965.

[20] Weigend A.S., Gershenfeld N.A. , Time series Prediction: Forecasting the Future and Understanding the Past, Reading, MA: Addison Wesley, 1994.

[21] Pachos, J.K. : Introduction to topological quantum computation, Univ. of Leeds, UK 2010.

[22] E. Artin,  Theory of Braids, Ann. of Math. (2) **48**: 101–126, 1947.